# Engineering Schottky Barrier in Black Phosphorus field effect devices for spintronic applications

*M.Venkata Kamalakar\*, B. N. Madhushankar, André Dankert, Saroj P. Dash[#]*

Department of Microtechnology .and Nanoscience, Chalmers University of Technology,
SE-41296, Göteborg, Sweden

e-mail: *venkata.mutta@chalmers.se; [#]saroj.dash@chalmers.se

**ABSTRACT**

Black phosphorous (BP) is recently unveiled as a promising two-dimensional direct bandgap semiconducting material. Here, we report the ambipolar field effect transistor behavior of multilayers of BP with ferromagnetic tunnel contacts. We observe a reduced Schottky barrier < 50 meV by using $TiO_2$/Co contacts, which could be further tuned by gate voltages. Eminently a good transistor performance is achieved in our devices, with drain current modulation of four to six orders of magnitude. The charge carrier mobility is found to be ∼155 and 0.18 $cm^2$ $V^{-1}$ $s^{-1}$ for holes and electrons respectively at room temperature. Furthermore, magnetoresistance calculations reveal that the resistances of the BP device with applied gate voltages are in the appropriate range for injection and detection of spin polarized holes. Our results demonstrate the prospect of engineering BP nanolayered devices for efficient nanoelectronic and spintronic applications.

# INTRODUCTION

The interest in two-dimensional (2D) van der Waals crystals and their heterostructures for next-generation electronics is rapidly growing since the past few years[1,2]. Over the last decade, graphene has attracted enormous attention for its extremely high carrier mobility[1,2] and long distance spin transport capabilites[3–5]. These qualities have large prospects for applications such as ultra-high frequency transistors[6] and spin communication[7]. However, pristine graphene lacks a band gap and spin-orbit coupling, which are crucial requirements for switching action of transistors[6] and spin based devices[7]. In this context, semiconducting 2D crystals of transition metal dichalcogenides (TMD) with larger bandgaps and spin-orbit coupling have been investigated intensively for future electronics devices[7–12]. However, TMD transistors show n-type behavior because of the atomic vacancies and the location of the charge neutral level in the vicinity of the conduction band. This considerably limits their role for CMOS applications, where both n and p-type transistor are required as complementary devices in logic circuits[7–9,11–13]. Additionally, spin transport in TMDs is largely eclipsed due to the intrinsically high spin-orbit coupling in them. Therefore, a 2D crystal possessing both charge and spin transport/control capability is necessary for integration of spin-based memory and logic operations in a single device.

The recently reported semiconducting 2D crystal "black phosphorous (BP)" possesses a direct bandgap of 0.3-2 eV depending on its thickness[14], and may lead to potential applications in nanoelectronics and optoelectronics. The field effect transistors (FET) and photo response have been demonstrated in BP nanolayers showing ambipolar property (both n and p-type) [14–18]. In addition to electronic transport, spin related properties of BP are also interesting, because of the low spin-orbit coupling and +1/2 nuclear spin of phosphorous (P) atoms[19]. This makes BP a unique platform for investigating the creation of spin polarized charge carriers[20], nuclear hyperfine interaction[21] and dynamical nuclear spin polarization[22] effects.

For these experiments, the realization of high performance BP transistors with ferromagnetic tunnel contacts constitute the primary requirements[20,23,24]. Such contacts can act as efficient spin injectors/detectors, and circumvent the conductivity mismatch issues[23,24]. However, it has been recently reported that there exists a Schottky barrier of 200 meV for holes at the Ti/BP interfaces[15]. With such energy barriers at the interface, only carriers that are thermally emitted over the Schottky barrier can flow across the contact, resulting in a reduced spin polarization and a low drive current[25–28]. Therefore, engineering of a reduced Schottky barrier via suitable contact materials is necessary to enhance the performance of BP transistor, which has not been explored so far[10,14]. In view of the interesting electronic and spintronics properties of BP, further investigations of this material are desirable to understand the electronic processes involved and unveil its prospect.

In this article, we present field effect transistor characteristics of multilayers of black phosphorous with ultra-thin $TiO_2$ and cobalt (Co) contacts. Using such ferromagnetic tunnel contacts, we demonstrate a reduction in Schottky barrier, which could further be tuned by gate voltages. The BP devices show ambipolar transistor behavior with efficient hole conduction, high on/off drain current and mobilities. Magnetoresistance calculations based on spin diffusion model[23,24] show that the transistor properties are in the appropriate range for exploration of spin related phenomena in BP.

## RESULTS AND DISCUSSIONS

The device scheme and an optical microscope image of a fabricated BP field effect transistor are presented in figure 1 (a) and (b) respectively. The BP flakes were prepared on a clean $SiO_2$ (285 nm), highly doped Si substrate. We used BP exfoliated from a bulk crystal (purchased from Smart Elements) using the conventional scotch tape method[2]. The flakes were identified by a combination of optical and atomic-force microscopy as shown in Supplementary Information Fig. S1. We used multilayer BP with a thickness of ~ 5 nm and widths of around ~ 3 µm. The $SiO_2$/n-Si is used as a back gate to control the carrier concentration in the BP. Contacts of $TiO_2$ (1nm)/Co (65 nm)/Al (2 nm) were fabricated on BP samples by electron-beam lithography, electron beam evaporation and lift-off techniques. Electrodes with widths 0.5–1 µm and a channel length of 1-3 µm were used. The BP flakes were covered by PMMA e-beam resist soon after exfoliation to avoid possible degradation upon longer exposure to air[16]. All the transport measurements were carried out under high vacuum conditions.

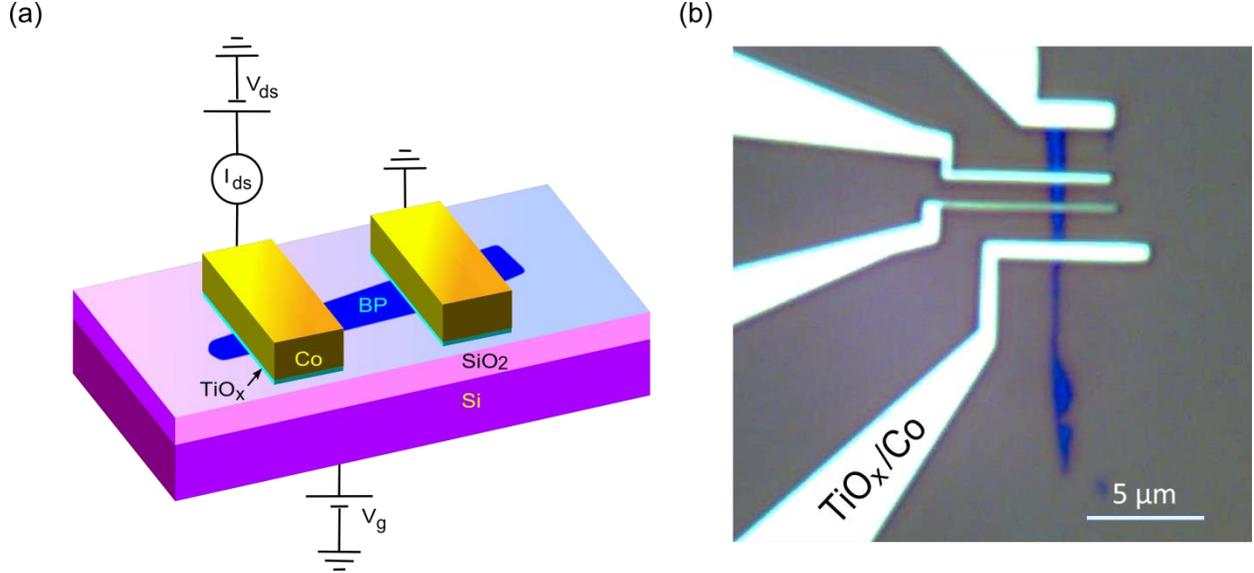

**Figure 1. Nanolayer black phosphorous (BP) field effect device** (a) Schematic of a BP FET with source, drain and back gate electrodes. (b) Optical image of a fabricated 5 nm BP FET with $TiO_2$ (1nm)/Co contacts and $SiO_2$/n-Si back gate.

To extract the Schottky barrier height at the Co/$TiO_2$/BP contacts, we performed detailed temperature dependent drain–source current ($I_{ds}$) voltage ($V_{ds}$) characteristics of device at different gate voltages. The BP devices showed ambipolar behavior with gate voltage ($V_g$), hole conduction for $V_g<0$ and electron conduction for $V_g>0$. In Fig.2a, we show representative temperature dependence of $I_{ds}$–$V_{ds}$ characteristics at gate voltage of $V_g= 20V$. Such dependence of $I_{ds}$ indicates the presence of a Schottky barrier at the interface. We analyze the data using the thermionic emission equation describing the electrical transport through a Schottky barrier into the BP channel [25–27]

$$I_{ds} = AA^*T^2 exp\left[-\frac{e}{k_BT}\left(\Phi_b - \frac{V_{ds}}{n}\right)\right] \quad (1)$$

where $A$ is the contact area, $A^*$ is the Richardson constant, $e$ is the electron charge, $k_B$ the Boltzmann constant, $\Phi_b$ is the Schottky barrier height, and $n$ is the ideality factor. Figure 2b shows the Arrhenius plot (ln ($I_{ds}T^{-2}$) versus $T^{-1}$) for different bias voltages $V_{ds}$. The slopes $S$ ($V_{ds}$) extracted from the Arrhenius plot follow a linear dependence with $V_{ds}$ ($S(V_{ds}) = -(e/k_B)(\Phi_b - (V_{ds}/n))$), as displayed in Figure 2c. The Schottky barrier is evaluated from the extrapolated value of the slope at zero $V_{ds}$ ($S_0 = -(e\Phi_b/k_B)$).

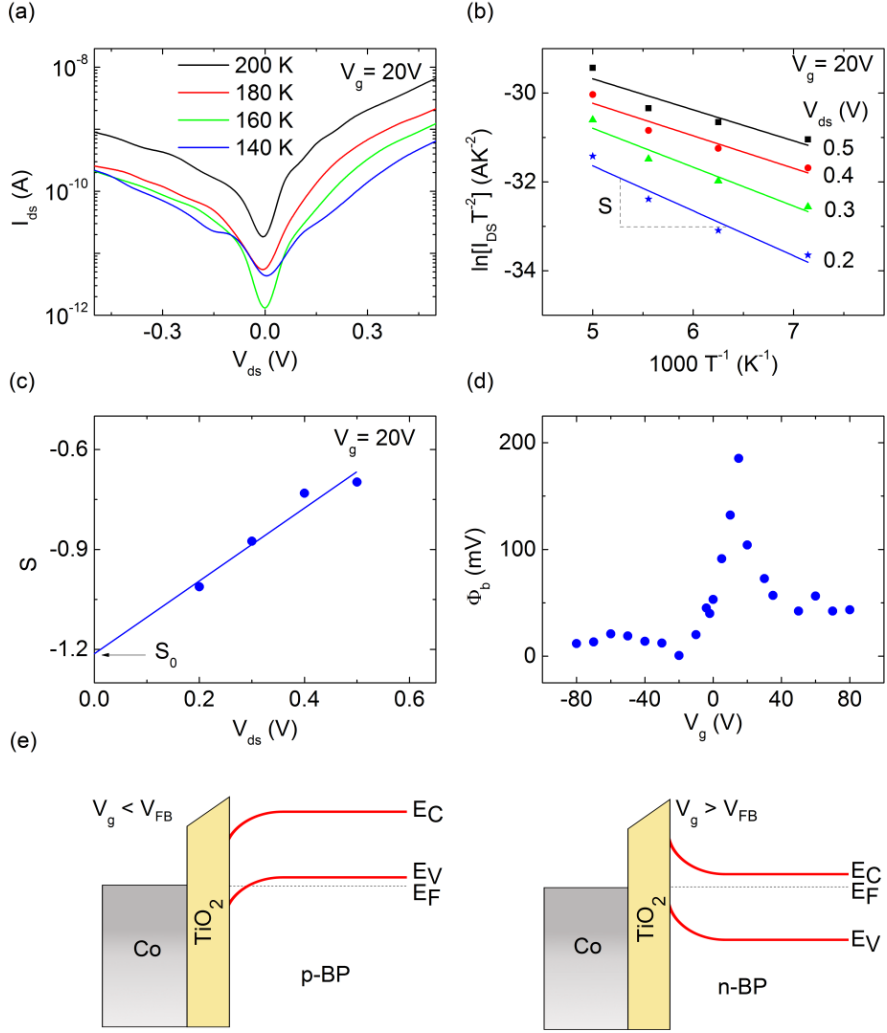

**Figure 2. Schottky barrier for BP/TiO$_2$/Co contact.** (a) Temperature dependence of $I_{ds}$–$V_{ds}$ at Vg=20V. (b) Arrhenius plot with $I_{ds}T^{-2}$ dependence on $T^{-1}$ with linear fits provides slope S at $V_g$=20 V and different $V_{ds}$. (c) Bias voltage dependence of the extracted slope $S$ provides the zero bias slope $S_0$, which is used to calculate the Schottky barrier height. (d) Schottky barrier height $\Phi_b$ extracted at different gate voltages. (e) Band diagram depicts the BP/TiO$_2$/Co interfaces for both p and n-type behavior of BP with gate voltage. $V_g$ and $V_{FB}$ represent gate voltage and flat band voltages respectively.

The evaluated Schottky barrier for different gate voltages is presented in Fig. 2d. For hole conduction ($V_g$ < 0), we observe a very small Schottky barrier of < 50 meV. Notably, the barrier heights obtained with our TiO$_2$/Co contacts are much lower than recent reports of 200 meV with Ti/Au contacts[14,15]. Higher work function of Co in comparison to Ti and/or de-pinning of Fermi level by ultra-thin TiO$_2$ may be responsible for such reduction of the Schottky barrier of BP in the hole conduction regime[10,25]. In the electron conduction conditions with $V_g > 0$, a higher Schottky barrier height $\Phi_b$ of 200 meV at $V_g$=20 V has been extracted. We note that the Schottky barrier can be tailored by applying a gate voltage. As shown in Fig.2d, the Schottky barrier height can be reduced to 25 meV for holes and 50 meV for electrons at high magnitudes of gate voltage. The lowering of the Schottky barrier at higher gate voltages for both electrons and holes are also evident from $I_{ds}$–$V_{ds}$ characteristics shown in Supplementary Information Fig. S2. Such lowering of Schottky barrier can be attributed to the tuning of band alignment of Co and BP and also doping of BP with gate voltage as depicted in Fig. 2e. The low Schottky barrier also gives rise to a lower contact resistance of $Co/TiO_2/BP$ devices. By applying a gate voltage, a drastic reduction of the contact $R_{cont}A = 1.4 \times 10^{-9}$ $\Omega m^2$ and a channel resistivity of $\rho_{ch} = 2.4 \times 10^{-4}$ $\Omega m$ could be achieved in the hole conduction regime. Similarly, for electrons the $R_{cont}A = 1.75 \times 10^{-9}$ $\Omega m^2$ and $\rho_{ch} = 7.2 \times 10^{-2}$ $\Omega m$ could be obtained. Such engineering of interface resistances using $Co/TiO_2$ contacts and gate voltages are of great significance for BP transistors with electrical spin injection capabilities.

Next we examine the field effect transistor (FET) characteristics of BP with $TiO_2/Co$ source and drain contacts and $SiO_2$/n-Si back gate. The output characteristics were obtained by measuring the drain–source current ($I_{ds}$) while sweeping the drain–source voltage ($V_{ds}$), at several gate voltages ($V_g$). Figure 3a presents the measured output characteristics showing a non-linear behavior at higher $V_{ds}$ and linear dependence only in a small bias voltage range. The presence of a finite Schottky barrier and the carrier depletion at the interfaces gives rise to such bias voltage dependence[10,14]. The measurements of $I_{ds}$–$V_{ds}$ at different gate voltages $V_g$ shows an efficient modulation of the drain current $I_{ds}$ and a transistor behavior at room temperature. Notably, significant source-drain conduction has been observed for both hole ($V_g$<0) and electron ($V_g$>0) conduction regime, proving ambipolar behavior of BP FET.

In Fig. 3b we show the room temperature transfer characteristics of BP transistor obtained by measuring the $I_{ds}$ with back gate voltage $V_g$ sweep with constant $V_{ds}$. The measured data shows a typical ambipolar BP FET with asymmetry in $I_{ds}$, which is determined by the position of the Co Fermi level close to the valence or conduction bands of the BP. The negative and positive gate voltages correspond to hole and electron doping in BP respectively. The high current densities reaching up to 150 µAµm$^{-2}$ for holes and 1 µAµm$^{-2}$ for electrons with carrier densities of 9 × $10^{12}$ cm$^{-2}$, make the device characteristics even more viable for applications. The ambipolar behavior in BP devices is dominant in p-type character, in contrast to the mostly observed n-type behavior of MoS$_2$ and WS$_2$ devices.

Our BP devices with ferromagnetic tunnel contacts show a reasonable transistor $I_{on}/I_{off}$ ratio with >$10^4$ for hole and >$10^2$ for electron conduction regimes at room temperature. We would like to emphasize here that it is possible to achieve even higher drain current modulation as the present on state current of our devices is below the saturation. Attaining such high values is limited by the achieved doping level by backgate voltage and breakdown electric field of the SiO$_2$ dielectric. By using high-k gate dielectrics, it should be possible to achieve higher carrier densities and hence a higher saturation of the drain current. We observe a subthreshold swing (SS) of ~ 11 V per decade (V/dec), which lies in the same order of magnitude as reported in other 2D semiconductor devices with 285 nm thick SiO$_2$ dielectric[10]. Compared to commercial Si devices with SS of 70 mV/dec, the larger SS obtained in our devices can be attributed to the thicker SiO$_2$ backgate dielectric and possible presence of interface states[27]. It is therefore expected that the SS can be significantly improved by using ultra-thin high-k dielectrics with minimum interface states.

The effective field-effect mobility in BP for both electrons and holes has been estimated from the measured transfer characteristics. We extract the slope in the linear region of the data by using the standard expression[8,10,14]

$$\mu = [(dI_{ds}) / (dV_g)] \times [L / (WC_iV_{ds})] \tag{2}$$

where *L= 690 nm* is the length and *W= 3.1* μm is the width of the channel, $V_{ds}$ = 0.5 V and $C_i$ is the capacitance between the channel and the back-gate per unit area ($C_i = \varepsilon_0\varepsilon_r/d$; $\varepsilon_r$= 3:9 for SiO$_2$; $d$ = 285 nm for back gate oxide thickness)[8]. Figure 3b shows the transfer curve, where $dI_{ds}/dV_g$ is the slope of the linear fit for mobility extraction. Since BP devices show an ambipolar behavior, we extract effective channel mobility for both electron and hole transport regimes. The extracted hole mobility is found to be 155 cm$^2$V$^{-1}$s$^{-1}$ for a 5 nm thick BP flake on SiO$_2$ substrates, which is comparable to reported values[15,17,18]. The mobility is also dependent on thickness of the BP, the highest value being reported for 10 nm flakes[15,17,18]. The electron mobility is extracted to be 0.18 cm$^2$V$^{-1}$s$^{-1}$, which is considerably lower than the hole mobility. The higher

Schottky barrier and lower mobility in electron conduction regime gives rise to three orders of magnitude lower current density than for holes. We have also observed similar mobility values for BP measured in four terminal configurations, which rules out the possible influence of the contacts (see Supplementary Information Fig. S3). The calculated mobilities may be the underestimated values as we have not considered the effects of anisotropic electronic properties in BP, defect states with back gate $SiO_2$ and possible surface degradation of BP even with short exposure to atmosphere[15,17,18]. We further examined the temperature dependence of the transistor characteristics of our BP FETs. The transfer curves measured at 2 and 300 K are shown in Fig. 3d. The transistor $I_{on}/I_{off}$ performance is increased by two orders of magnitude at 2 K. These results clearly indicate that BP field effect transistors with ferromagnetic tunnel contacts, show a high mobility and significantly large values of on state current for hole conduction regime. Such p-type BP transistors can be used in complementary circuits together with n-type $MoS_2$ transistors for 2D transparent electronics.

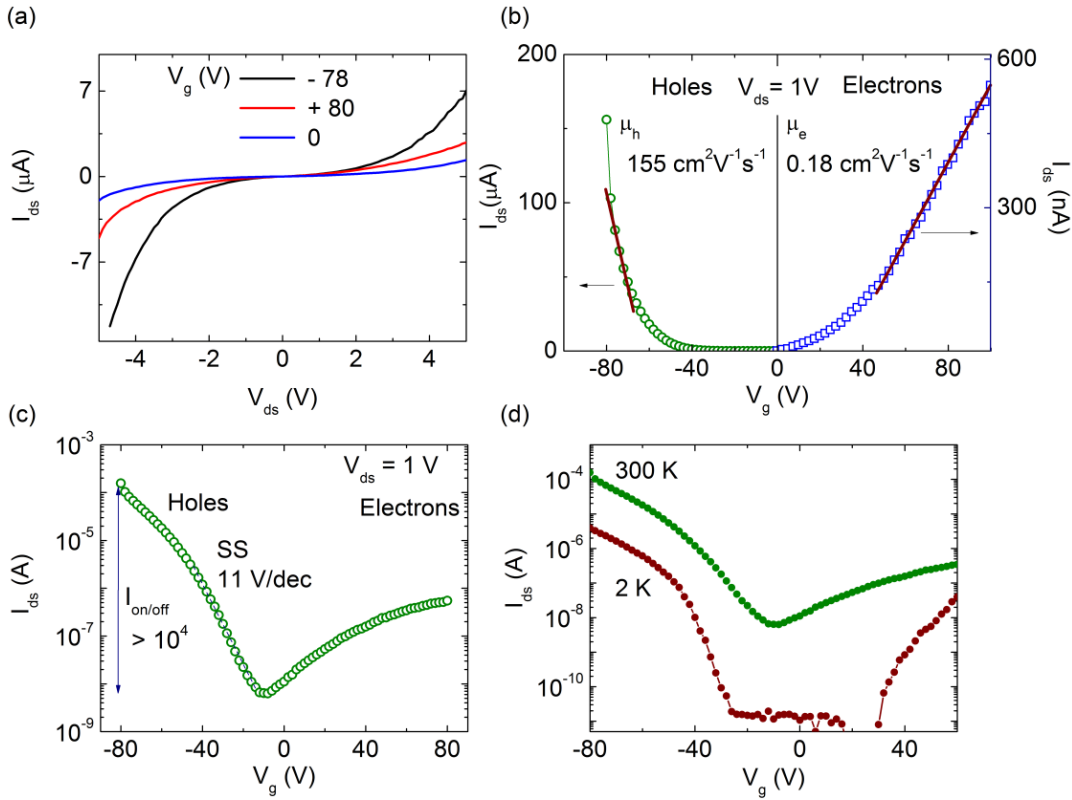

**Figure 3. Characterization of BP FETs.** (a) Output characteristics $I_{ds}$–$V_{ds}$ measured in high bias range for different gate voltages ($V_g$). (b) Transfer characteristic in linear scale for $V_{ds} = 1$V. The effective field-effect mobility for electrons and holes are 155 and 0.18 cm$^2$ V$^{-1}$ s$^{-1}$ respectively, as calculated from the subthreshold slope of the transfer curve. (c) Transfer

characteristic in logarithmical scale for $V_{ds} = 1V$ with an $I_{on}/I_{off} > 10^4$ and sub-threshold swing SS ~ 11 V/dec. (d) Temperature dependence of BP transfer characteristic $I_{ds}$–$V_g$ for $V_{ds}= 1V$ measured at 2 and 300 K.

In addition to the electronic properties of p-type BP devices, spin transport studies of holes are also interesting because of the low spin-orbit coupling in BP and the presence of nuclear spin of ½ on each of it's P atom[19]. A versatile way of creating spin polarization in semiconductors is the electrical spin injection from ferromagnetic tunnel contacts[20]. However, the resistances of the contacts and semiconductors channels have to be in the optimum range to observe magnetoresistance effects. In order to grasp the capability of BP for spin polarized transport, we evaluate the contact and channel resistances on the basis of spin diffusion theory[23]. We consider a FM/I/BP/I/FM structure, where I is the interface contact and FM is the ferromagnet. The optimal contact and channel resistances required are estimated by considering experimentally measured mobility and resistance values. To observe a high MR, the interface resistance should lie within the range

$$\rho_N L < R_{cont} A < \rho_N \frac{l_{sf}^2}{L} \qquad (3)$$

where $\rho_N$ is the resistivity of the BP channel of length L, $R_{cont}$ is the contact resistance with a contact area A, $l_{sf}$ is the spin diffusion length. The calculated MR as a function of contact *RA* is shown in Fig. 4a, where we have considered $L = 50$ nm and different expected spin lifetimes (0.01- 1 ns) for holes in BP. Experimentally, lateral L of < 50 nm can be achieved by electron beam lithography, shadow evaporation[29] or using vertical device structures of BP. The spin lifetime for holes in the valence band are expected to be in the above mentioned range and not exactly known at this stage.. In BP,Spin relaxation mechanism due to nuclear hyperfine interaction may exit[21,22]. However, experiemnts performed on nuclear spin of C-13 atomic graphene do not show significant influence on spin lifetime. Thus, the adopted spin lifetime values are reasonable considering the low spin orbit coupling expected in BP. The MR calculations show the possibility of observation of a significant magnetoresistance in BP in a narrow contact resistance range. The low MR for small contact resistances is due to the conductivity mismatch, preventing a spin injection from the Co into BP. For larger contact resistances, the reduction in MR is caused by the relaxation of the spins in the BP channel due to longer dwell time[23].

The contact resistance is $R_{cont}A = 1.4 \times 10^{-9}$ $\Omega m^2$ for on state of BP FET in the hole conduction regime with $V_g$=-80 V, and lies in the optimum range for the observation of a large MR. In fig. 4b, we also show that the calculated MR as a function of gate voltage $V_g$ and contact resistance $R_{cont}A$ for hole conduction. The observed change in MR with gate voltage is because of the tuning of the hole carrier density in the BP channel by the gate voltage. The contact resistance of our device is shown by the dashed lines in Figure 4a and 4b. Our results demonstrate the tunability of the channel and contact resistances, and expected MR by applying a gate voltage. The TiO$_2$/Co spin tunnel contacts are in the optimum range for spin injection and detection in BP in higher hole doping ranges. Although our present devices with TiO$_2$/Co contacts show more favorable conditions for spin polarized hole injection, similar results for electrons spin could be achieved by using lower work function materials at the interfaces[30–32]. Such design will align the Fermi level near the conduction band edge and lower contact resistances to more optimal values[32]. Further enhancements in the MR can be achieved by improving the mobility of the BP channel by high k-dielectrics encapsulation[8]. These heterostructures of BP with ferromagnetic contacts will allow exploring electrical methods[20,33,34] to create spin polarization in BP, and also alternative approaches such as thermal[35] and dynamical[36] spin injection.

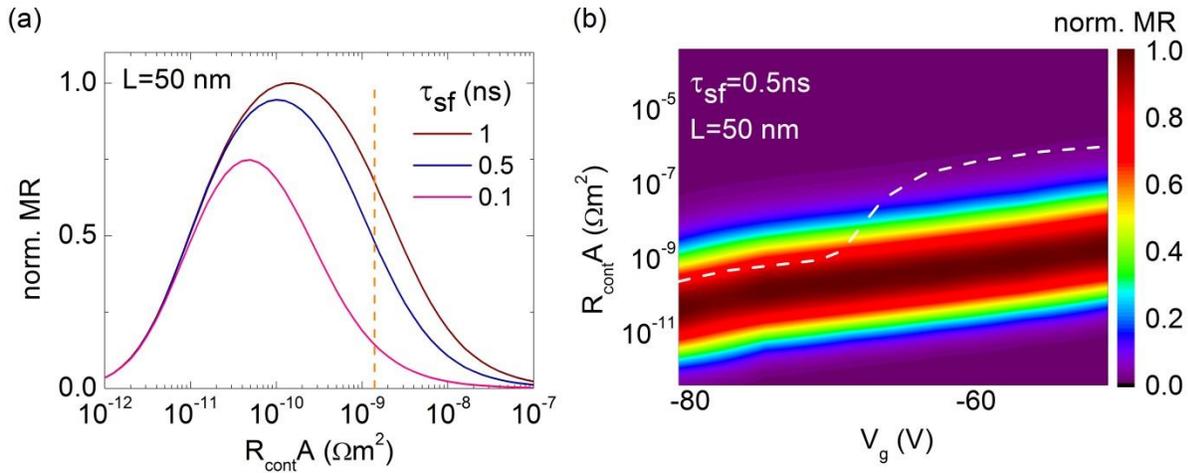

**Figure 4. Magnetoresistance (MR) calculation** – Calculated MR of FM/I/BP/I/FM spin FET structure as a function of the contact resistance area product ($R_{cont}A$) for a channel length L= 50 nm at room temperature. The values are normalized to the peak value at a spin lifetime $\tau_{sf}$ = 1 ns. (a) MR calculation for different spin lifetimes $\tau_{sf}$ at a gate voltage $V_g$ = - 80 V (on-state). The dashed lines represent the $R_{cont}A$ value for the TiO$_2$/Co tunnel contact on BP in our experiments. (b) MR as a function of gate voltage and contact resistance for $\tau_{sf}$=0.5 ns and L=50nm. The dashed lines represent the variation of contact $R_{cont}A$ of our devices with gate voltage.

## CONCLUSION

In conclusion, we have investigated the field effect transistor characteristics of nanolayers of black phosphorous with ferromagnetic tunnel contacts. Lower contact resistances and ambipolar behavior with both hole and electron conduction could be achieved by using $TiO_2$/Co contacts and gate voltage. The Schotkky barrier heights could be reduced to values less than 50 mV for both holes and electrons. We demonstrate ambipolar transistor with $I_{on}/I_{off} > 10^4$ and on-state current reaching 150 µA/µm in the hole conduction regime. At low temperatures the transistor characteristics show an increase in $I_{on}/I_{off}$ by two orders of magnitude $> 10^6$, which are comparable to other 2D semiconducting crystal based devices. Due to a smaller bandgap of multi-layer BP devices, both electron and hole conduction could be achieved by application of gate voltages. The effective field effect channel mobility for holes and electrons are found to be ~155 and 0.18 $cm^2$ $V^{-1}$ $s^{-1}$ respectively at room temperature. Our calculations for the observation of magnetoresistance in BP indicate that the resistances of BP/$TiO_2$/Co FET can be tuned to the optimum range by the application of gate voltage. The integration of such ferromagnetic tunnel contacts on BP circumvents the conductivity mismatch problem with promising transistor performance. This opens up the prospect for spin based nanoelectronic devices using 2D semiconducting crystals.

**Acknowledgements**

The authors acknowledge the support from colleagues of Quantum Device Physics Laboratory and Nanofabrication Laboratory at Chalmers University of Technology. This research is financially supported by Nano Area of Advance program at Chalmers University of technology, EU FP7 Marie Curie Career Integration grant, and Swedish Research Council Young Researcher Grant.

# Supplementary Information

# Engineering Schottky Barrier in Black Phosphorus field effect devices for spintronic applications

*M. Venkata Kamalakar\*, B. N. Madhushankar, André Dankert, Saroj P. Dash[#]*

Department of Microtechnology and Nanoscience, Chalmers University of Technology,
SE-41296, Göteborg, Sweden

e-mail: \*venkata.mutta@chalmers.se; [#]saroj.dash@chalmers.se

**Nanolayers of Black Phosphorous:** Flakes of black phosphorous are exfoliated onto $Si/SiO_2$ substrate by the scotch tape technique. Flakes having thicknesses ~ 5 nm are selected by a combination of optical and atomic force microscopy (Fig. S1).

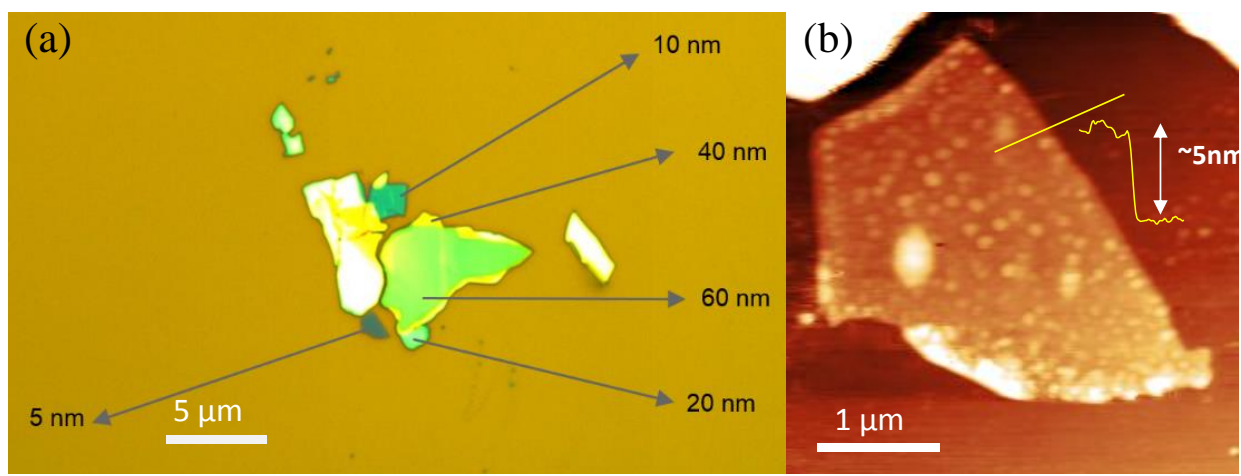

Figure S1: (a) Optical microscope image of a group of Black phosphorus flakes with various thicknesses. (b) Atomic force microscope image of a 5 nm thin flake.

**Schottky Barrier for the electron and hole regimes:** Reduction in Schottky barrier in the electron and hole regimes is evident from the weak temperature dependence of the $I_{ds}$–$V_{ds}$ characteristics obtained on the device at the two extremes of applied gate voltage as presented in Fig. S2

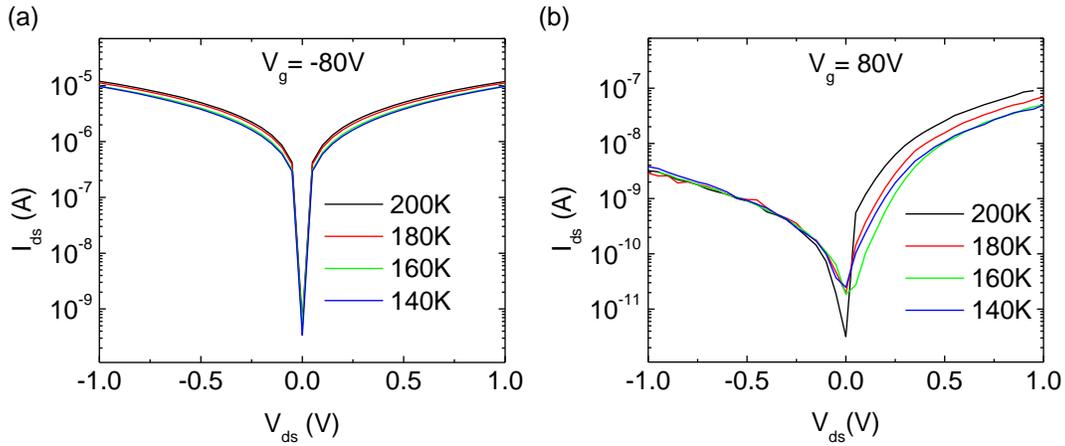

Figure S2: $I_{ds}$-$V_{ds}$ characteristics at different temperatures for the (a) Hole regime at $V_g$= - 80 V (b) Electron regime at $V_g$= 80 V.

**Mobility extraction from four probe measurements:** In order to rule out the possible influence of contacts on the evaluation of field effect mobility, we performed the gate dependent four probe measurements of the channel. The hole and electron mobilities using the four probe measurements were found to be close to the values obtained in FET measurements.

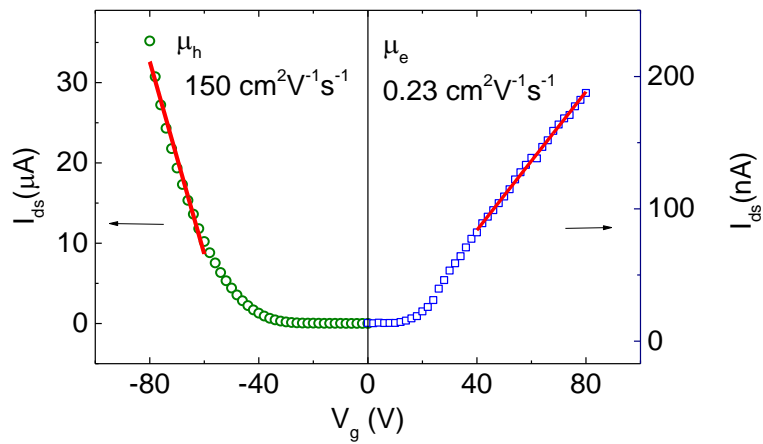

Figure S3: Gate dependent four probe measurements of the BP channel with extraction of hole and electron motilities at room temperature.